\documentclass[aps,prl,floatfix,superscriptaddress,noshowpacs,twocolumn]{revtex4-2}
\usepackage{graphics,graphicx,epsfig,amsmath,amssymb,epstopdf,wasysym,comment}
\usepackage{hyperref}
\usepackage[dvipsnames]{xcolor}
\usepackage{braket,amsfonts}
\usepackage{dsfont}
\usepackage{hyperref}
\usepackage[dvipsnames]{xcolor}
\usepackage{amsthm}

\newcommand{\be}{\begin{equation}}
\newcommand{\ee}{\end{equation}}
\newcommand{\bga}{\begin{gather}}
\newcommand{\ega}{\end{gather}}
\newcommand{\bea}{\begin{eqnarray}}
\newcommand{\eea}{\end{eqnarray}}

\newcommand{\bK}{\mathbf{K}}

\newcommand{\bq}{\mathbf{q}}

\newcommand{\bk}{\mathbf{k}}

\newcommand{\bkp}{\mathbf{k'}}
\newcommand{\bp}{\mathbf{p}}

\newcommand{\br}{\mathbf{r}}
\newcommand{\Ima}{\text{Im}}
\newcommand{\Rea}{\text{Re}}
\newcommand{\dis}{\displaystyle}

\newcommand{\up}{\uparrow}
\newcommand{\down}{\downarrow}
\newcommand{\fract}[2]{\frac{\dis \;#1\;}{\dis \;#2\;}}
\newcommand{\Tr}{\mathrm{Tr}}
\newcommand{\eqn}[1]{(\ref{#1})}

\newcommand{\ep}{{\epsilon}}

\newcommand{\bw}{\begin{widetext}}
\newcommand{\ew}{\end{widetext}}
\newenvironment{eqs}%
{\begin{equation} \begin{aligned}}%
{\end{aligned} \end{equation} }
\newcommand{\beal}{\begin{eqs}}
\newcommand{\eal}{\end{eqs}}
\newcommand{\bd}[1]{{\boldsymbol{#1}}}
\newcommand{\esp}[1]{\text{e}^{#1}}
\newcommand{\bealn}{\beal\nonumber}

\begin{document}

\title{Spin-liquid insulators can be Landau's Fermi liquids}

\author{Michele Fabrizio}
\affiliation{International School for Advanced Studies (SISSA), Via Bonomea 265, I-34136 Trieste, Italy} 

\begin{abstract}
The long search for insulating materials that possess low-energy quasiparticles carrying electron's quantum numbers except charge  
-- inspired by the neutral spin-1/2 excitations, the so-called spinons, exhibited by Anderson's resonating-valence-bond state -- seems to have reached a turning point after the discovery of several Mott insulators displaying 
same thermal and magnetic properties as metals, including quantum oscillations in a magnetic field. Here, we show that such anomalous behaviour is not inconsistent with Landau's Fermi liquid theory of quasiparticles at a Luttinger surface. 
That is the manifold of zeros within the Brillouin zone of the single-particle Green's function at zero frequency, and which thus defines the spinon Fermi surface conjectured by Anderson.  

\end{abstract}

\maketitle

Common sense would suggest that Mott insulators and Landau's Fermi liquids are antinomic phases of matter that can turn one into the other 
only through a Mott transition. \\
However, there is growing, intriguing
evidence of \textit{quasiparticle}-like excitations in some Mott insulating materials.  
For instance, the Kondo insulators SmB$_6$ and YbB$_{12}$ show quantum 
oscillations in a magnetic field, finite specific heat, $C_v/T$, and 
thermal conductivity, $\kappa/T$, coefficients for $T\to 0$ 
~\cite{SmB6-Suchitra-Science2015,SmB6-Suchitra-NatPhys2018,YbB12-Xiang-Science2018,YbB12-Sato-NatPhys2019,SMB6-Suchitra-iScience2020,YbB12-Xiang-PRX2022}, though $\kappa\sim T$ is still debated in SmB$_6$~\cite{SMB6-Xu-PRL2016,SmB6-Boulanger-PRB2018}. \\
Evidence of finite $C_v/T$ and $\kappa/T$ for $T\to 0$ is also found in candidate spin-liquid insulators:   
1$T$-TaS$_2$~\cite{Ribak-PRB2017,Yu-PRB2017,Murayama-PRR2020}, 
and, with some caveats, in the organic salts
EtMe$_3$Sb[Pd(dmit)$_2$]$_2$~\cite{Matsuda-Science2010,Yamashita-NatComm2011,Matsuda-NatComm2012,Ni-PRL2019,Taillefer-PRX2019,Yamashita-SciRep2022} 
and 
$\kappa$-(BEDT-TTF)$_2$Cu$_2$(CN)$_3$~\cite{Kanoda-NatPhys2008,Yuji-NatPhys2009}. 
Quantum oscillations in the magnetothermal conductivity of the field induced spin-liquid state of $\alpha$-RuCl$_3$ have also been reported~\cite{RuCl3-1}, even though their origin is controversial~\cite{RuCl3-2}. 

All the above properties, at odds with the conventional view of insulators, are 
commonly interpreted by the existence of \textit{neutral quasiparticles}
~\cite{Lee-PRL2005,Lee-PRL2007,Motrunich-PRB2006,Lee-Science2008,Lee-PNAS2017,SmB6-Senthil-NatCom2018,Senthil-PRB2018}, not necessarily gapless~\cite{Lee-PNAS2017}, although alternative explanations 
have been proposed~\cite{Cooper-PRL2015,Zhang-PRL2016,Cooper-PRL2017}. 
Those quasiparticles are dubbed \textit{spinons}~\cite{PWA-RVB,PWA-PRL1987} when they only carry the spin quantum number, 
which is the case of systems whose low energy behaviour is determined by just a single band, as we 
shall
assume hereafter. 

Despite the observed Fermi-liquid-like thermal and magnetic properties of spinons, their emergence from  spin-charge deconfinement~\cite{Baskaran-SSC1987} is 
at first sight incompatible with Landau's Fermi liquid theory~\cite{Landau1,Nozieres&Luttinger-1,Nozieres&Luttinger-2}.  This is obviously the 
case of conventional Landau's quasiparticles at a Fermi surface, the location of poles of the single-particle Green's function 
at zero frequency and temperature, since 
these poles 
entail metallicity. \\
However, it has been recently shown~\cite{mio-2} 
that Landau's quasiparticles also exist at a Luttinger surface, the manifold of zeros of the single-particle Green's function 
at zero frequency and temperature. These quasiparticles are invisible in the single-particle spectrum, and 
are also 
incompressible~\cite{mio-3}, thus perfectly allowed  
in insulators. Nonetheless, the insulating character poses  
constraints to Landau's Fermi liquid theory, most notably 
the
vanishing 
of
Drude weight and 
of
charge compressibility. Here, we show 
that these constraints can be fulfilled.  
We conclude
that a Landau Fermi liquid 
can well be insulating, and analyse its physical properties with special emphasis on the quantum oscillations in a magnetic field. \\


\noindent
\textit{Uncovering Landau quasiparticles --}
In what follows, we consider a periodic model 
 with a single band of interacting electrons, and assume that neither translational 
symmetry nor spin rotational one are broken. \\
The single-particle Green's function is therefore diagonal in momentum $\bk$ 
and spin $\sigma=\up,\down$, and independent of the latter.
In Matsubara frequencies, $\ep=(2n+1)\pi T$, the Green's function satisfies Dyson's equation
\beal
G(i\ep,\bk) &= \fract{1}{\;i\ep-\ep(\bk)-\Sigma(i\ep,\bk)\;}\;,
\label{Dyson}
\eal
where $\ep(\bk)$ is the non-interacting energy dispersion in momentum space measured with respect to the chemical 
potential, and $\Sigma(i\ep,\bk)$ the self-energy that, like $G(i\ep,\bk)$, 
has a real part even in $\ep$, while 
\beal
\Ima\,\Sigma(i\ep,\bk) = -\Ima\,\Sigma(-i\ep,\bk) 
\begin{cases}
<0 & \ep >0\,,\\
>0 & \ep <0\,.
\end{cases}
\label{Im Sigma}
\eal
We define the real function 
\beal
Z(\ep,\bk) &= Z(-\ep,\bk)= \left(1-\fract{\;\Ima\,\Sigma(i\ep,\bk)\;}{\ep}\right)^{-1}\,,
\label{Z}
\eal
which, because of \eqn{Im Sigma}, varies in the interval $[0,1]$. Through $Z(\ep,\bk)$ we 
can rewrite Eq.~\eqn{Dyson} as 
\beal
G(i\ep,\bk) &= \fract{Z(\ep,\bk)}{\;i\ep-\ep_*(\ep,\bk)\;}\;,
\label{Dyson-qp}
\eal
with real 
\beal
\ep_*(\ep,\bk)=\ep_*(-\ep,\bk) = Z(\ep,\bk)\,\Big(\ep(\bk)+\Rea\,\Sigma(i\ep,\bk)\Big)\,.
\label{ep_*}
\eal
Landau's Fermi liquid theory 
can be formally derived under the assumption that $\ep_*(\ep,\bk)$ and $Z(\ep,\bk)$ are analytic, at least to leading order, in $\ep$  
around $\ep=0$, as well as in $\bk$ close to the surface defined by 
$\ep_*(0,\bk)=0$~\cite{mio-2}. This assumption is equivalent to assuming that 
$\Sigma(i\ep,\bk)$ is analytic at any non-zero $\ep$, which includes conventional Fermi liquids as the special case 
of $\Sigma(i\ep,\bk)$ analytic also at $\ep=0$, but also allows for poles of $\Sigma(i\ep,\bk)$ for $\ep\to 0$. 

The actual \textit{quasiparticles} have energy dispersion $\ep_*(\bk)\equiv\ep_*(0,\bk)$ and  
residue $Z(\bk)\equiv Z(0,\bk)$. The roots of $\ep_*(\bk)$ in momentum space define the \textit{quasiparticle Fermi surface} that, 
because of the definition \eqn{ep_*}, correspond 
\begin{itemize}
\item either to the roots of $\ep(\bk)+\Rea\,\Sigma(0,\bk)$, the conventional 
Fermi surface, 
\item or those of $Z(0,\bk)$, the so-called Luttinger surface~\cite{Igor-PRB2003}.
\end{itemize}
Therefore, well-defined quasiparticles exist 
at Fermi as well at Luttinger surfaces, and that despite the vanishing 
quasiparticle residue $Z(\bk)$ at the Luttinger surface implies the absence of 
quasiparticle peaks in the physical electron density of states.\\

\noindent
\textit{Fermi liquid properties --} 
We recall that Landau's Fermi liquid theory allows calculating linear response functions at low temperature, low frequency 
and long wavelength in terms of two unknown functions:
the quasiparticle dispersion $\ep_*(\bk)$ and the Landau parameters 
$f_{\bk\sigma,\bkp\sigma'}$, where $\sigma$ and $\sigma'$ are the spins of the quasiparticles with momentum $\bk$ and $\bkp$, respectively.
In reality, this huge simplification just applies to densities of conserved quantities and their currents defined through the 
continuity equation. Indeed, only in those cases one can exploit the Ward-Takahashi identities and relate vertex to self-energy corrections~\cite{Nozieres&Luttinger-1}.
\\
In a single-band periodic model, the conserved quantities are 
the electron number $N=N_\up+N_\down$, the energy $E$, and the magnetisation along a given axis, e.g., $M=N_\up-N_\down$. 
We denote by $\chi_{\rho_Q}(\omega,\bq)$ and 
$\chi^{\phantom{q}}_{J_Q}(\omega,\bq)$, the proper response 
functions, respectively, of the density, $\rho_Q$, and current, $J_Q$, operators associated to 
the conserved quantity $Q=N,E,M$, i.e., the response functions irreducible with respect to cutting a Coulomb interaction line. 
The thermodynamic susceptibilities are simply obtainable through $\chi_Q = - \chi^q_{\rho_Q}$, 
where $\chi^q_{\rho_Q}\equiv \chi_{\rho_Q}(\omega=0,\bq\to\bd{0})$ is the so-called 
$q$-limit of the density response function. We recall that the specific heat is actually defined through $C_v=\chi_E/T$. \\
In absence of impurities, the low-temperature conductivities have the standard Drude-like expression 
$\sigma_Q(\omega) = i\,D_Q/(\omega+i0^+)$, 
where the Drude weights $D_Q$ coincide with the 
so-called $\omega$-limit of the corresponding current response functions: $D_Q=\chi^\omega_{J_Q}\equiv 
\chi^{\phantom{q}}_{J_Q}(\omega\to0,\bq=\bd{0})$. Similarly to the specific heat, the thermal conductivity is 
defined by $\sigma_E(\omega)/T$. \\

\noindent
According to Landau's Fermi-liquid theory~\cite{Nozieres&Luttinger-1,Nozieres&Luttinger-2}
\beal
\chi_{N/M} &= -2\!\int \fract{d\bk}{(2\pi)^d}\, 
\fract{\partial f\big(\ep_*(\bk)\big)}{\partial\ep_*(\bk)}\,\Big(1-\text{A}_{S/A}(\bk)\Big)\,,\\
D_{N/M} &= -\fract{2}{d}\!\int \fract{d\bk}{(2\pi)^d}\,  
\fract{\partial f\big(\ep_*(\bk)\big)}{\partial\ep_*(\bk)}\;\bd{v}_*(\bk)\cdot
\bd{v}_{S/A}(\bk)\,,
\label{compressibility and susceptibility}
\eal
where $d>1$ is the dimension (in $d=1$ Landau's Fermi liquid theory is 
not applicable~\cite{Solyom}), $f(x)$ the Fermi distribution function, 
$\bd{v}_*(\bk)=\partial\ep_*(\bk)/\partial\bk$ the quasiparticle group velocity, and 
\beal
\text{A}_{S/A}(\bk) &= - \int \fract{d\bkp}{(2\pi)^d}\, 
\fract{\partial f\big(\ep_*(\bkp)\big)}{\partial\ep_*(\bkp)}\;\text{A}^{S/A}_{\bk,\bkp}\,,\\
\overline{\bd{v}}_{S/A}(\bk) &= \bd{v}_*(\bk) + 
 \int \fract{d\bkp}{(2\pi)^d}\, 
\fract{\partial f\big(\ep_*(\bkp)\big)}{\partial\ep_*(\bkp)}\;\bd{v}_*(\bkp)\,f^{S/A}_{\bk,\bkp}\,.
\label{definitions}
\eal
The parameters $\text{A}^{S/A}_{\bk,\bkp}$ correspond to the $q$-limit of the quasiparticle scattering 
amplitudes in the spin-singlet ($S$) and spin-triplet ($A$) 
particle-hole channels, and are related to the $f$-parameters, the 
$\omega$-limit counterparts, 
\beal
f_{\text{S}\,\bk,\bkp} &= f_{\bk\up,\bkp\up} + f_{\bk\up,\bkp\down}\,,\\
f_{\text{A}\,\bk,\bkp} &= f_{\bk\up,\bkp\up}- f_{\bk\up,\bkp\down}\,,
\label{fS and fA}
\eal
through the Bethe-Salpeter equation 
\bealn
\text{A}^{S/A}_{\bk,\bkp} = f^{S/A}_{\bk,\bkp}+\int \fract{d\bp}{(2\pi)^d}\,\fract{\partial f\big(\ep_*(\bp)\big)}{\partial\ep_*(\bp)}
f^{S/A}_{\bk,\bp}\,
\text{A}^{S/A}_{\bp,\bkp}\,.
\eal
Similarly, the specific heat $C_v$ and the Drude weight $K$ of the thermal conductivity read
\bealn
C_v &= -\fract{2}{T}\,\int \fract{d\bk}{(2\pi)^d}\fract{\partial f\big(\ep_*(\bk)\big)}{\partial\ep_*(\bk)}\,\ep_*(\bk)^2\\
&\quad -\fract{2}{T}\,\int \fract{d\bk \,d\bkp}{(2\pi)^{2d}}
 \fract{\partial f\big(\ep_*(\bk)\big)}{\partial\ep_*(\bk)}\,
\fract{\partial f\big(\ep_*(\bk')\big)}{\partial\ep_*(\bk')}\\
&\qquad\qquad\qquad\qquad\qquad \ep_*(\bk)\,\ep_*(\bkp)\;
\text{A}^S_{\bk,\bkp}\,,\\
K&= -\fract{2}{dT}\,\int \fract{d\bk}{(2\pi)^d} \fract{\partial f\big(\ep_*(\bk)\big)}{\partial\ep_*(\bk)}\,\ep_*(\bk)^2\,
\big|\bd{v}_*(\bk)\big|^2\\
&\qquad +\fract{2}{dT}\,\int \fract{d\bk \,d\bkp}{(2\pi)^{2d}}\fract{\partial f\big(\ep_*(\bk)\big)}{\partial\ep_*(\bk)}\,
\fract{\partial f\big(\ep_*(\bk')\big)}{\partial\ep_*(\bk')}\\
&\qquad\qquad \qquad \qquad 
\ep_*(\bk)\,\ep_*(\bkp)\,\bd{v}_*(\bk)\cdot\bd{v}_*(\bkp)\;
f^\text{S}_{\bk,\bkp}\,.
\eal
The first term on the right hand side of both equations is linear in temperature $T$. 
Conversely, the second terms give a finite contribution at low $T$ only upon expanding 
$\text{A}^S_{\bk,\bkp}$ and $f^\text{S}_{\bk,\bkp}$ in 
$\ep_*(\bk)$ and $\ep_*(\bkp)$, as well as including higher order corrections in the heat vertex 
as obtained through the Ward-Takahashi identity. All those corrections yield at first sight 
terms of order $T^3$. In reality, 
the expansion is not regular. For instance, the corrections to the linear term of the specific heat are actually 
of order $T^d$~\cite{Chubukov-PRB2005,Chubukov-PRB2006}, 
with logarithmic corrections in $d=3$, $T^3\,\ln1/T$. Nonetheless, at leading 
order in $T$ only the first terms contribute, and thus 
\beal
C_v &\simeq \fract{2\pi^2}{3}\,T\,\rho_*\,,&
K &\simeq  C_v\, \fract{v_*^2}{d}\,,
\label{heat 2}
\eal
where 
\beal
\rho_* \equiv \int \fract{d\bk}{(2\pi)^d}\,\delta\big(\ep_*(\bk)\big)\,,
\label{DOS}
\eal
is the quasiparticle density of states at the chemical potential, and 
\beal
v_*^2 \equiv \fract{1}{\rho_*}\, 
\int \fract{d\bk}{(2\pi)^d}\,\delta\big(\ep_*(\bk)\big)\,
\big|\bd{v}_*(\bk)\big|^2\,.
\label{average v}
\eal\\

\noindent
\textit{Mott insulators with a Luttinger surface --} 
Let us now consider a hypothetical model that has only a Luttinger surface
in the Brillouin zone, with finite quasiparticle density of states 
at the chemical potential, $\rho_*\not=0$ in Eq.~\eqn{DOS}. 
Since quasiparticles at the Luttinger surface are invisible in the single-particle density of states and 
incompressible~\cite{mio-3}, the system describes a non-symmetry breaking Mott insulator that may only occur at half-filling in a single-band model.\\
In a Mott insulator with localised electrons, we expect that $f_{\bk\up,\bkp\up}\simeq 0$, which implies $f^S_{\bk,\bkp}\simeq - f^A_{\bk,\bkp}$ 
and $\text{A}^S_{\bk,\bkp}\simeq - \text{A}^A_{\bk,\bkp}$.  
However, for the system to be a charge insulator, we need to impose that the compressibility $\chi_N$ and charge Drude weight $D_N$ in Eq.~\eqn{compressibility and susceptibility} vanish, which implies, through Eq.~\eqn{definitions},   
that $\text{A}_S(\bk)= 1$ plus a correction that averages to zero on the Luttinger surface, 
as well as that the flux of $\bd{v}_{S}(\bk)$ out of the Luttinger surface is zero. 
In turn, since $\text{A}_A(\bk)\simeq -\text{A}_S(\bk)= -1$ and 
$\bd{v}_{A}(\bk) \simeq 2\bd{v}_*(\bk) - \bd{v}_{S}(\bk)$, 
then, through Eqs.~\eqn{compressibility and susceptibility} and \eqn{average v}, the  
spin susceptibility $\chi_{M}$ and Drude weight $D_{M}$ 
become simply
\beal
\chi_{M} &\simeq 4\rho_*\,,&
D_{M} &\simeq \fract{4}{d}\,\rho_*\,v_*^2\,.
\label{spin-Mott}
\eal
Comparing \eqn{spin-Mott} with \eqn{heat 2}, we 
find that the Wilson ratio, which measures the effective correlation strength, is
\beal
R_\text{W} = \fract{\pi^2 T}{3 C_v}\;\chi_M\simeq 2\,.
\eal
Therefore, a Landau Fermi liquid characterised by a Luttinger surface without 
Fermi pockets may indeed have charge properties of an insulator, while spin and thermal ones of 
a metal, in that not dissimilar from a spin-liquid 
insulator with gapless spinons.\\
We mention that conventional Fermi liquids often do not survive down to $T=0$, since they may encounter an instability at $T_c>0$ towards a different phase that, most of the times, breaks symmetries and opens gaps in the quasiparticle spectrum. Well known examples are the superconducting 
and superfluidity instabilities in normal metals and $^3$He, respectively. 
A Fermi liquid description of such an instability is justified when 
quasiparticles have already reached quantum degeneracy at $T_c$, which implies 
that $T_c$ must be much smaller than the quasiparticle Fermi energy $\ep_F$. \\
Similarly, we cannot exclude that also quasiparticles at a Luttinger surface, the gapless spinons, may become unstable at $T_c\ll \ep_F$ towards, e.g., a magnetically ordered phase, and eventually acquire a gap. In this case, which 
presumably corresponds to highly frustrated magnets, 
the above Fermi liquid properties would still be observable for $T_c\ll T\ll \ep_F$. On the contrary, if $T_c\sim \ep_F$, likely the case of unfrustrated 
magnets, the quantum degenerate behaviour of quasiparticles at the Luttinger surface cannot set in before the instability. \\

\noindent
\textit{Quantum oscillations --} 
The next relevant question to be addressed is whether quasiparticles at a Luttinger surface 
contribute to quantum oscillations in a magnetic field $B$. On one hand, the semiclassical approach to the de Haas-van Alphen (dAvH) effect 
by Lifshitz and Kosevich~\cite{Lifshitz}, which just relies on the existence 
of quasiparticles, would suggest a positive answer. However, the vanishing Drude weight implies, through \eqn{compressibility and susceptibility} and \eqn{definitions}, that  
\bealn
0 &= -\int d\bk\, \fract{\partial f\big(\ep_*(\bk)\big)}{\partial \bk}
\cdot \bd{v}_S(\bk)\\
&= \int d\bk\, f\big(\ep_*(\bk)\big)\,\bd{\nabla}_\bk\cdot \bd{v}_S(\bk)
\\
&=\int d\bk\, f\big(\ep_*(\bk)\big)\,\Tr\Big(\hat{m}_c(\bk)^{-1}\Big)\,,
\eal
where $\hat{m}_c(\bk)$ is the cyclotron mass tensor as it emerges from the Landau-Boltzmann transport equation. 
Considering, for simplicity, an isotropic $\hat{m}_c(\bk) = m_c(\bk)\,\hat{I}$, it follows that vanishing Drude weight is equivalent to vanishing  $1/m_c(\bk)$, or, equivalently, vanishing cyclotron frequency, once integrated over the volume enclosed by the Luttinger surface. That hints at the absence of quantum oscillations, in contrast to the previous observation.

To resolve this issue, we resort to Luttinger's theory of the de Haas-van Alphen effect in 
interacting electron systems~\cite{Luttinger-dHvA}. Luttinger showed that the leading oscillatory part of the free energy 
derives from
\beal
\Delta F_\text{osc} &= -T\,\sum_\ep\, \esp{i\ep 0^+}\,\Tr \ln\Big(i\ep-\hat{H}_0 -\hat{\Sigma}(i\ep)\Big)\,,
\label{F_osc}
\eal
where $\hat{H}_0$ is the non-interacting Hamiltonian, which includes the static and uniform magnetic field $B$, 
represented in a generic basis of single particle wavefunctions. The self-energy matrix $\hat{\Sigma}(i\ep)$ in \eqn{F_osc} must include 
any polynomial in $B$ but not oscillatory terms in $1/B$~\cite{Luttinger-dHvA}. In matrix notations, we now define
\bealn
\hat{Z}(\ep)^{-1} &\equiv 1 -\fract{\;\Ima\,\hat{\Sigma}(\ep)\:}{\ep}\;,
\eal
which is a positive-definite matrix with eigenvalues $\geq 1$, and 
the hermitian matrix 
\bealn
\hat{H}_*(\ep) &= \sqrt{\,\hat{Z}(\ep)\;}\,\Big(\hat{H}_0 +\Rea\,\hat{\Sigma}(i\ep)\Big)\,
\sqrt{\,\hat{Z}(\ep)\;}\,.
\eal
With these definitions that generalise \eqn{Z} and \eqn{ep_*}, the free energy component 
\eqn{F_osc} becomes
\beal
\Delta F_\text{osc} &= -T\,\sum_\ep\, \esp{i\ep 0^+}\,\Tr \ln\Big(i\ep-\hat{H}_*(\ep) \Big)
\\
&\qquad +T\, \sum_\ep\, \esp{i\ep 0^+}\,\Tr\ln \hat{Z}(\ep)
\\
&\equiv \Delta F^{(1)}_\text{osc} + \Delta F^{(2)}_\text{osc}\,.
\label{F}
\eal
In conventional Fermi liquids, where $\hat{Z}(0)$ has no null eigenvalue,
the first term, $\Delta F^{(1)}_\text{osc}$, is the only that contributes and yields the Lifshitz and Kosevich theory of the dHvA effect, 
as shown by Luttinger~\cite{Luttinger-dHvA}. Indeed, in the semiclassical limit, $\hat{H}_*(\ep)$ becomes 
the representation in the chosen basis of the operator $\ep_*\big(\ep,\bK(\br)\big)$, Eq.~\eqn{ep_*} with  
$\bk$ replaced by 
\beal
\bK(\br) = -i\hbar\,\fract{\partial}{\partial\br} + \fract{e}{2c}\,\bd{B}\wedge\br\,,
\eal
and thus
\beal
\Delta F^{(1)}_\text{osc} &\simeq 
-T\sum_\ep\,\esp{i\ep 0^+}\;\Tr\ln\Big(i\ep-\ep_*\big(\bK(\br)\big)\Big)\,.
\label{F1}
\eal
After that, one can simply follow Lifshitz and Kosevich~\cite{Lifshitz} and derive the expression of the dHvA oscillations. 
\\
However, in the present case of a Luttinger surface, also $\Delta F^{(2)}_\text{osc}$  in \eqn{F} may contribute since 
$\hat{Z}(\ep)$ has zero eigenvalues 
at $\ep=0$.  To assess their role, we note that $\hat{Z}(\ep)$ in the semiclassical limit is the representation of the operator 
$Z\big(\ep,\bK(\br)\big)$, i.e., of $Z(\ep,\bk)$ in Eq.~\eqn{Z} with $\bk\to \bK(\br)$. Moreover, the contribution of $\Delta F^{(2)}_\text{osc}$ to 
quantum oscillations only derives from the region around the zeros 
of $Z(\ep,\bk)$~\cite{Wasserman-PRB1979}, i.e., small $\ep$ and 
$\bk$ close to the Luttinger surface. In that region, we can write, 
without loss of generality and consistently with the analytic assumption, 
that~\cite{Rice-RPP2011,*Alexei-RPP2019,mio-2} 
\beal
\Sigma(i\ep,\bk)\underset{\ep\to 0}{\simeq} \fract{\Delta(\bk)^2}{\;
i\ep-E(\bk)\;}\;,
\label{ansatz}
\eal
where $\bk_L:\,E(\bk_L)=0$ defines the Luttinger surface provided $\Delta(\bk_L)\not=0$, so that, for $\ep\simeq 0$ and $\bk\simeq \bk_L$, 
\beal
Z(\ep,\bk) &= \fract{\ep^2+E(\bk)^2}{\;\ep^2+E(\bk)^2+\Delta(\bk)^2\;}\\
&\simeq \fract{\;\ep^2+E(\bk)^2\;}{\;\Delta(\bk)^2\;}\;,\\
\ep_*(\ep,\bk) &= \fract{\;
\ep(\bk)\,\big(\ep^2+E(\bk)^2\big) - E(\bk)\,\Delta(\bk)^2\;}
{\;\ep^2+E(\bk)^2+\Delta(\bk)^2\;}\\
&\simeq -E(\bk)\,,
\label{ansatz-1}
\eal
which, as anticipated, are analytic. 
Therefore,  
\bealn
&Z\big(\ep,\bK(\br)\big)\simeq \ep^2 + \ep_*\big(\bK(\br)\big)^2 \\
&\qquad\qquad  = \Big(i\ep-\ep_*\big(\bK(\br)\big)\Big)\, \Big(-i\ep-\ep_*\big(\bK(\br)\big)\Big)\,,
\eal
and, correspondingly, 
\beal
\Delta F^{(2)}_\text{osc} &\simeq 
T\sum_\ep\,\esp{i\ep 0^+}\;\bigg[\ln\Big(i\ep-\ep_*\big(\bK(\br)\big)\Big)\\
&\qquad\qquad\qquad \quad
+\ln\Big(-i\ep-\ep_*\big(\bK(\br)\big)\Big)\bigg]\,,
\label{F2}
\eal
so that, through \eqn{F1} and \eqn{F2}, Eq.~\eqn{F} becomes 
\beal
\Delta F_\text{osc}&\simeq 
T\sum_\ep\,\esp{i\ep 0^+}\;
\ln\Big(-i\ep-\ep_*\big(\bK(\br)\big)\Big)\\
&\simeq -\Delta F^{(1)}_\text{osc}\,,
\label{F-final}
\eal
as can be readily verified following Lifshitz and Kosevich~\cite{Lifshitz}. 
As a result, quasiparticles at the Luttinger surface of a Mott insulator 
do yield dHvA oscillations in the magnetisation 
$-\partial\Delta F_\text{osc}/\partial B$ alike conventional quasiparticles with 
dispersion $\ep_*(\bk)$, apart from a $\pi$-shift.
\\

\noindent
\textit{Concluding remarks --} 
Few remarks are now in order. 
Conventional theories of spin-liquids~\cite{Kivelson-PRB1987,Kalmeyer-PRL1987,Read&Sachdev-PRL1989,Wen-PRB2002,Senthil-PRL2003,Lee-RMP2006,Ng-RMP2017} predict that a spinon Fermi surface is most likely associated to so-called 
$U(1)$ spin liquids, apart from few known exceptions~\cite{Yao-PRL2009,Baskaran-arxiv2009,Orthogonal-metals,Elio-PRB2020}. In that $U(1)$-case, 
the specific heat behaves at low temperature as $T^{2/3}$ and $T\ln1/T$ 
in $d=2$ and $d=3$, respectively~\cite{Lee-RMP2006,Senthil-PRB2008,Sachdev-RMP2022}, 
and, correspondingly, $\kappa/T$ diverges for $T\to 0$~\cite{Lee-PRL2005}.  
These thermal properties, different from the observed ones, challenge   
the spin-liquid interpretation. Finite $C_v/T$ and $\kappa/T$ for $T\to 0$ 
may be, for instance, attributed to magnetic impurities, assuming a gapped spin liquid phase lacking a spinon Fermi surface~\cite{Lee-PNAS2017}. However, this 
explanation implies that also quantum oscillations are not due to spinons, and thus 
that all intriguing thermal and magnetic properties observed in experiments are 
unrelated to the purported spin liquid nature of the material, which is a bit disappointing. \\ 
On the contrary, the Fermi liquid properties of a Mott insulator 
with a Luttinger surface seem to account for all experimental evidences. 
Nonetheless, the analyticity assumption on the self-energy underlying Landau's Fermi liquid theory is evidently incompatible with the above mentioned non-analytic behaviour of $U(1)$ spin liquids with a spinon Fermi surface.  
Therefore, either that analytic behaviour never 
occurs in physical models, or Mott insulators with a Luttinger surface 
realise one of the above mentioned exceptions~\cite{Yao-PRL2009,Baskaran-arxiv2009,Orthogonal-metals,Elio-PRB2020} of spin liquids with a spinon Fermi surface.\\ 
Indeed, an example of a spin liquid with $C_v\sim T$ is very well known: the half-filled Hubbard model in one dimension. 
Even though interacting electrons in $d=1$ behave as 
Luttinger liquids~\cite{Haldane-LL-1981}, their 
low-frequency, low-temperature and long-wavelength properties are just alike 
conventional Fermi liquids~\cite{D&L,Solyom,Haldane-LL-1981}, including the specific heat that, as we mentioned, is obtainable by the $q$-limit of the heat-heat response function. 
In particular, the half-filled Hubbard model in $d=1$ is an insulator that 
has a Luttinger surface at $k=\pm\pi/2$ as well as gapless spinons that yield a finite spin susceptibility, a finite $C_v/T$, apart from corrections vanishing as powers of
$1/\ln T$, and a Wilson ratio $R_W=2$ for $T\to 0$~\cite{Troyer-PRB2000}. That is precisely what our Fermi-liquid analysis predicts.

In conclusion, we have shown that non-symmetry breaking Mott insulators with a Luttinger surface realise gapless spin liquids, where the spinons are actually Landau's quasiparticles at the Luttinger surface, which thus 
provides the rigorous definition of Anderson's spinon Fermi surface~\cite{PWA-RVB,PWA-PRL1987}. These quasiparticles contribute to thermal and magnetic properties, including quantum oscillations, just like conventional quasiparticles do, despite the system is a charge insulator. \\

\noindent
The author is very grateful to Andrey Chubukov and Erio Tosatti for helpful discussions and comments. This work was funded by the European Research Council (ERC), 
under the European Union's Horizon 2020 research and
innovation programme, Grant agreement No.~692670 "FIRSTORM".

\bibliographystyle{apsrev4-2}
%

\end{document}